\documentclass[global,referee]{svjour}

\usepackage{color}
\usepackage{graphicx}
\usepackage{amsmath}
\usepackage{amssymb}
\usepackage{cite}
\newcommand{\modif}[1]{#1}
\newcommand{\modb}[1]{#1}
\newcommand{\moda}[1]{#1}
\newcommand{\modc}[1]{#1}
\begin{document}


\title{Locking the frequency of lasers to \moda{an optical cavity at the $1.6 \times 10^{-17}$ relative} instability level}


\author{Qun-Feng Chen \and Alexander Nevsky \and Stephan Schiller}

\institute{Institut f\"{u}r Experimentalphysik, Heinrich-Heine-Universit\"{a}t D\"{u}sseldorf, 40225 D\"{u}sseldorf, Germany
}
\date{}
\maketitle
\begin{abstract}
  We stabilized the frequencies of two independent Nd:YAG lasers to two adjacent longitudinal modes of a high-finesse Fabry-P\'erot resonator and obtained a beat frequency instability of 6.3 mHz at an integration time of 40 s. Referred to a single laser, this is $1.6\times10^{-17}$ relative to the laser frequency, and $1.3\times10^{-6}$ relative to the full width at half maximum of the cavity resonance. The amplitude spectrum of the beat signal had a FWHM of 7.8 mHz. This stable frequency locking is of importance for next-generation optical clock interrogation lasers and fundamental physics tests.
\end{abstract}



 \noindent 
Ultrastable lasers are key tools for precision metrology, e.g. in optical clocks \cite{Takamoto:2005:321,RevModPhys.83.331,Margolis:2009:}, laser interferometric gravitational-wave observatories \cite{Abbott:2009:076901}, and optical resonator-based tests of Lorentz invariance \cite{PhysRevLett.103.090401,RevModPhys.83.11}. Currently, the lowest absolute frequency instabilities achieved \moda{by ultrastable lasers alone} are at the level of 0.8 to $2\times 10^{-16}$ \cite{Jiang:2011:158,2011arXiv1112.3854K}, determined by comparing two independent lasers each locked to a resonator. Theoretical studies have shown that thermal noise of the resonators is a fundamental limit in laser frequency stabilization \cite{PhysRevLett.93.250602,PhysRevA.73.031804}, and current experimental performance is at that level. The thermal instability limit depends on a number of resonator parameters and various promising approaches are being followed to lower it. One of these is operation at cryogenic temperature \cite{PhysRevLett.78.4741,Muller:2003:020401,2011arXiv1112.3854K}. In order to take advantage of the expected reduction in the resonator thermal limit, the frequency stabilization technique itself must have a correspondingly low instability. 

In this paper we report on the achievement of a frequency locking instability compatible with the thermal noise in next-generation resonators. Our study is based on the well-known approach of locking two lasers to two different modes (frequencies $\nu_{1}$, $\nu_{2}$) of the resonator, with a frequency difference $\nu_{1} - \nu_{2}$ on the order of one free spectral range \cite{Salomon:88,DAY:1992:1106,Ruoso:1997:259,vonZanthier:2005:1021}. The influence of changes of resonator length on the frequency difference is reduced by several orders of magnitude compared to the influence on each frequency alone. We chose equal transverse modes (TEM$_{00}$), differing in longitudinal mode number by 1. It is then also expected that the thermal noise effect on the modes is nearly equal, as the fluctuations of the mirrors' surfaces should affect in the same way the cavity modes that have almost identical electric field patterns. We used a room-temperature high-finesse ULE resonator to which we locked two monolithic Nd:YAG lasers by means of the Pound-Drever-Hall (PDH) technique, and achieved a lock instability as low as 5 mHz for a single laser. This result is about 5 times better than the best previous results \cite{Salomon:88,Ruoso:1997:259,vonZanthier:2005:1021} for integration times on the order of 100 s.

\begin{figure}[tb]
  \begin{center}
	\includegraphics[width=0.8\textwidth]{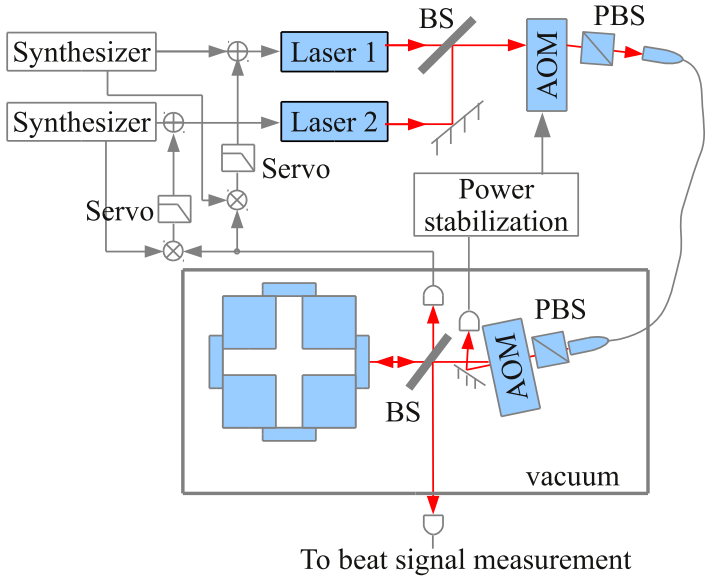}
  \end{center}
  \caption{Schematic setup. BS: non-polarizing beam splitter; PBS: polarizing beam splitter; AOM: acousto-optic modulator; servo: lock circuit for frequency stabilization; $\otimes$: mixer; $\oplus$: bias-T, respectively. For clarity, only the fast frequency control channels, to the piezoelectric transducers on each laser, are shown.}
  \label{fig:setup}
\end{figure}

The setup of the experiment is shown in Fig. \ref{fig:setup}. The lasers are nonplanar monolithic Nd:YAG lasers (1064 nm). The 8.4 cm long cavity is one of the cavities in the double-cavity ULE block used previously in a test of Lorentz Invariance \cite{Eisele20081189,PhysRevLett.103.090401} upgraded with fused silica mirror substrates and ULE rings for thermal expansion coefficient modification \cite{Legero:10}. The free spectral range (FSR) and linewidth of the cavity are 1.78 GHz and 3.4 kHz, respectively, corresponding to a finesse of about 520 000. The cavity is temperature stabilized to about 15 $^\circ$C, close to the temperature of zero coefficient of thermal expansion. The two laser beams are combined in air by using a non-polarizing beam splitter (BS) and are then guided into a vacuum chamber by a 1.5 m long, non-polarization-maintaining, single-mode, APC optical fiber with a vacuum feed-through. Inside the vacuum chamber, the waves are coupled to the TEM$_{00}$ modes of the resonator with a total coupling efficiency of around 85\%. An acousto-optic modulator (AOM) in air after the BS serves as an isolator. A second AOM inside the vacuum chamber is used to suppress interferences between the optical fiber and the resonator. The undiffracted output of the vacuum AOM is monitored by a photodetector and the signal serves to stabilize the total optical power available before the resonator by controlling the RF power of the AOM outside the chamber. The two polarizing beam splitters (PBS) before and after the optical fiber are used to set the polarization of the lasers. The BS between the vacuum AOM and the cavity sends a part of the input wave to a photodetector where the beat note between the two lasers is generated, and a part of the waves reflected from the cavity to a photodetector for PDH signal detection. The laser powers on the detector are about 60 $\mu$W for each laser.

\moda{The optical setup used in this experiment was designed to be relatively insensitive to the surroundings. First, the essential part of the optical setup is built inside a vacuum chamber, which eliminates the influences of air, whose fluctuations influence the optical length, the beam position, and the residual amplitude modulation (RAM) on the detector. Second, the optical paths of the two lasers are identical between the cavity and the beat signal detector, which decreases the requirement of the stability of the optical length by a factor of $\nu/f$, where $\nu$ is the optical frequency of the laser and $f$ is the frequency of the beat signal [19]. Third, to reduce \modif{the influence of} the deformation of the evacuated vacuum chamber by the pressure of the surrounding atmosphere, the internal optical components are mounted on a breadboard placed \modif{inside} the chamber. Copper mesh is used to decouple the breadboard \modb{mechanically} from the chamber while \modif{providing} a thermal link between them. This approach ensures that the optical setup will not be misaligned upon evacuation of the chamber.}

\begin{figure}[tb]
  \begin{center}
	\includegraphics[width=0.8\textwidth]{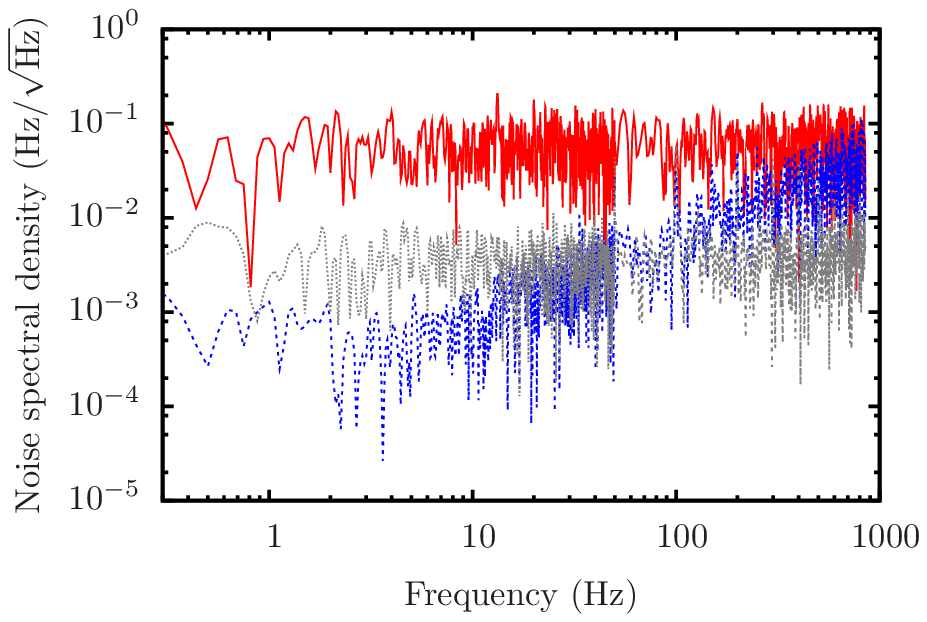}
  \end{center}
  \caption{Frequency noise spectral densities. The values are obtained by measuring the voltage noises at the error signal monitor outputs of the lock circuit, and scaling with the sensitivities (slopes) of the error signal. The measurement was performed using a network signal analyzer (SR780) with 800 frequency bins. Red solid line: the laser is off resonance with the cavity. This noise level \moda{(65 mHz/$\sqrt{\rm Hz}$)} shows the combined noise of the laser, the detector, and the lock circuit input stage noise (before the monitor output). Gray dotted line: lasers are blocked. This noise \moda{(3.7 mHz/$\sqrt{\rm Hz}$)} is due to the electronic noise of the detector and of the lock circuit. Blue dashed line: the laser frequency is in lock. The noise spectra of the both laser systems are similar.}
  \label{fig:noise}
\end{figure}

The standard PDH technique is used to stabilize the frequencies of the lasers. The phase modulations are applied directly to the lasers via piezoelectric transducers (PT) pressing on the laser resonators. The modulation frequencies are 2.958 MHz and 2.213 MHz for laser 1 and laser 2, respectively. The two error signals are obtained from the same photodetector. Its output is split into two parts, each of which is mixed with a modulation signal to produce the respective error signals. The error signals are fed back to the PTs of the lasers for fast control and to the heaters of the laser resonators for slow control. We use two-stage PID circuits in the locking electronics\moda{. The gain is proportional to $f^{-2}$ up to a frequency of about 3 kHz (defined by the cavity linewidth)}, and constant gain for larger frequencies $f$ up to 100 kHz. The closed-loop gains are around $1\times 10^{8}$ at 1 Hz. The laser locking bandwidths of about 30 kHz is limited by the first strong electro-mechanical resonance of the PTs. Figure \ref{fig:noise} displays the suppression of the noise by the locking system and demonstrates that the main noise in the system is from the laser. The laser intensity noise and the electronic noise will be transferred to the frequency noise of each laser when its frequency is locked; therefore, these noises lead to a lower limit for the short-term frequency instability of the lock.
  
Because the two laser waves propagate through the same optical fiber with the same polarization and are coupled to the same spatial mode of the cavity, the noises introduced by the optical fiber and the thermal noise of the cavity are nearly common-mode. The frequencies of the lasers differ by one free spectral range of the cavity (1.78 GHz). The beat signal of the two lasers is mixed down to about 50 kHz by using a synthesizer (Agilent E8241A) set to 1 782 824 600 Hz. The beat signal is measured by a frequency counter and the data is record by a PC. In order to realize a dead-time-free frequency counter, we used two standard frequency counters (HP 53181A) that alternately measure and transfer the data to the PC. The frequency reference inputs of all instruments are connected a GPS-stabilized crystal reference generator. 

\begin{figure}[tb]
  \begin{center}
	\includegraphics[width=0.4\textwidth]{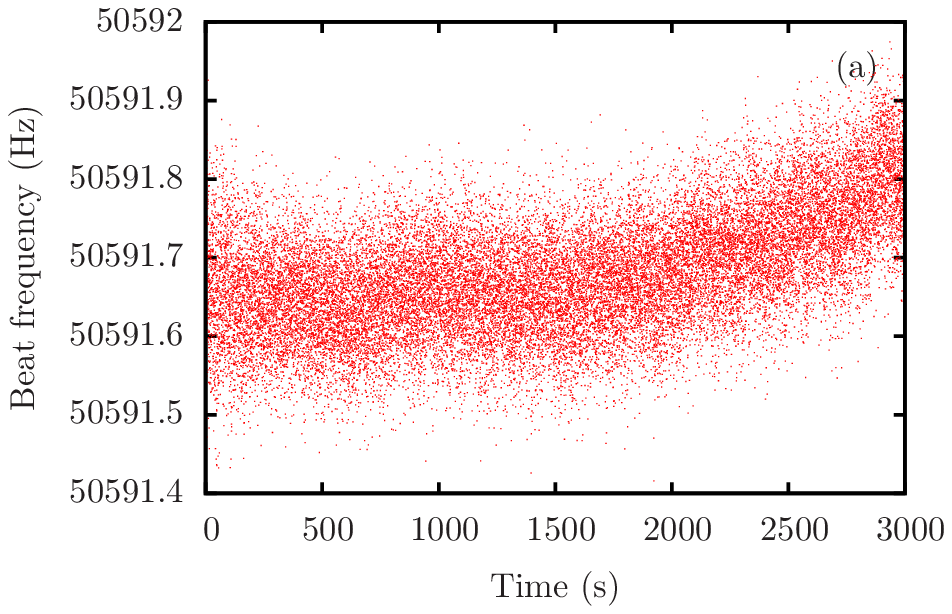}~~\includegraphics[width=0.4\textwidth]{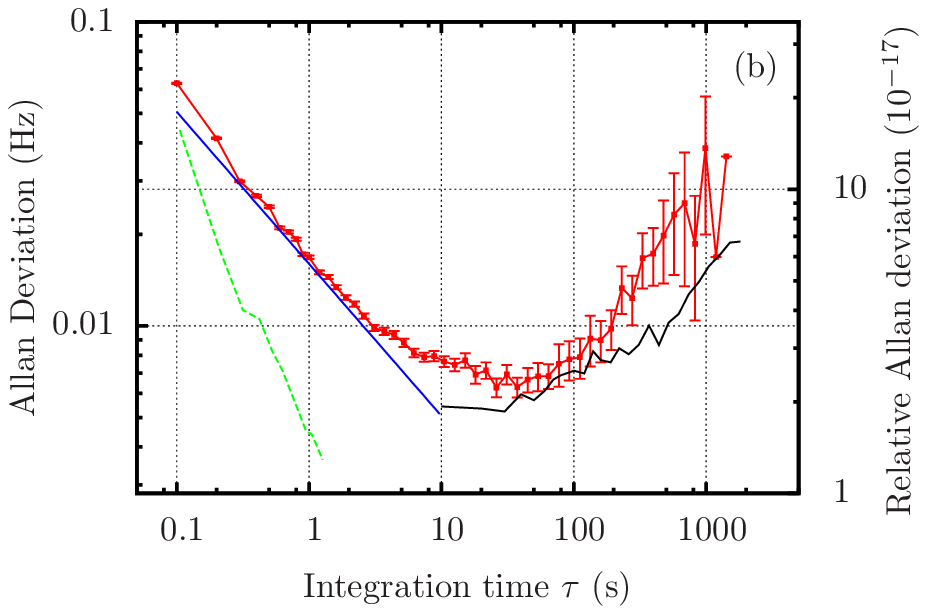}
  \end{center}
  \caption{(a) the frequency of the beat signal. (b) the corresponding Allan deviation. The blue line is the estimated contribution from the intensity noise of the lasers, the detector, and the lock circuits, assuming the noise levels are proportional to the laser powers. The green dashed line is the contribution from the noise of the synthesizer and of the frequency measuring system. The black line is the estimated contribution from the instability of RAM strength and the mixer.}
  \label{fig:result1}
\end{figure}

A 3000 s long segment of the obtained beat frequency data and the corresponding Allan deviation are shown in Fig. \ref{fig:result1}. The minimum Allan deviation is $2.2\times 10^{-17}$ (6.3 mHz) at an integration time of 40 s, which corresponds to a lock instability for an individual laser of $1.6\times 10^{-17}$ assuming the two systems being very similar. 
   
The limit imposed by the laser noise and the frequency locking system can be estimated from the noise spectral densities shown in Fig. \ref{fig:noise} (red line), because these noises will be transferred to the frequency noises of the lasers when the laser frequencies are locked. \moda{Figure \ref{fig:noise} (red line) shows that the noise spectral density is nearly frequency-independent (white noise), with a value of about 65 mHz/$\sqrt{\rm Hz}$. From this value, a lower limit for the Allan deviation of the beat frequency can be derived. 
In order to estimate the noise spectral density under \modc{lock, where the power on the detector decreased by about a factor of 4,} the contributions to the noise were analyzed: (i), electronic noise \modif{(3.7 mHz/$\sqrt{\rm Hz}$, gray dotted line in Fig. \ref{fig:noise}) was} much smaller than the total value (65 mHz/$\sqrt{\rm Hz}$), (ii) the (theoretical) shot noise of the laser at 15 $\mu$W \modif{contributed} a frequency noise of only 0.6 mHz/$\sqrt{\rm Hz}$. Therefore, the \modb{total level (red line in Fig. \ref{fig:noise})} is mainly \modb{given by} the excess noise of the laser and, \modb{furthermore, it is} expected to be proportional to the laser power, $P$.  Assuming this relationship, the lower limit of the Allan deviation is estimated to be 16 mHz $(\tau/1 \textrm{ s})^{-1/2}$, which is indicated in Fig. \ref{fig:result1} (b) as the blue line.} This value agrees well with the measured Allan deviation of the beat signal ($17.4\pm0.3$ mHz at 1 s). This result indicates that our locking quality is limited by the excess noise of the laser, the major contribution to the red line in Fig. \ref{fig:noise}. \moda{The relatively large intensity noise in our experiment \modif{was} due to the choice of the relatively low modulation frequency (ca. 3 MHz), which \modif{was} not sufficiently far away from the relaxation oscillation frequency of the laser, ca. 400 kHz.}

We also checked the limitation due to the frequency measuring system (FMS), excluding the contributions of the lasers and the frequency locking systems. It was estimated by replacing the beat signal of the lasers with another synthesizer (Agilent E8241A) which was phase-locked to a hydrogen maser. The obtained Allan deviation is shown as the green dashed line in Fig. \ref{fig:result1} (b). At the integration time of 0.1 s, the instability of the FMS is close to the Allan deviation derived from the noise level (blue line in Fig. \ref{fig:result1} (b)), which suggests that our result is limited by the instability of the FMS at short-time scale ($\lesssim 0.1$ s). 

\begin{figure}[tb]
  \begin{center}
	\includegraphics[width=0.4\textwidth]{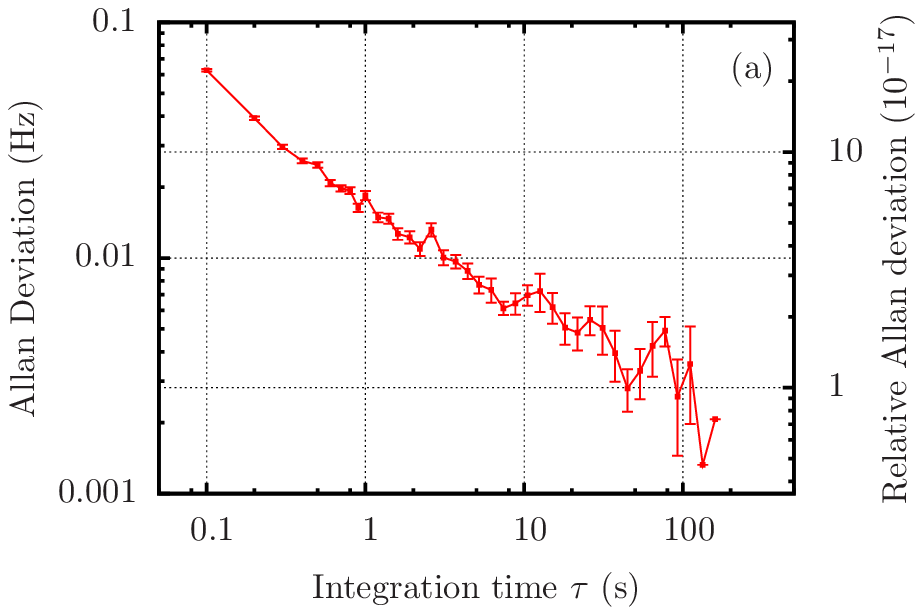}~~\includegraphics[width=0.4\textwidth]{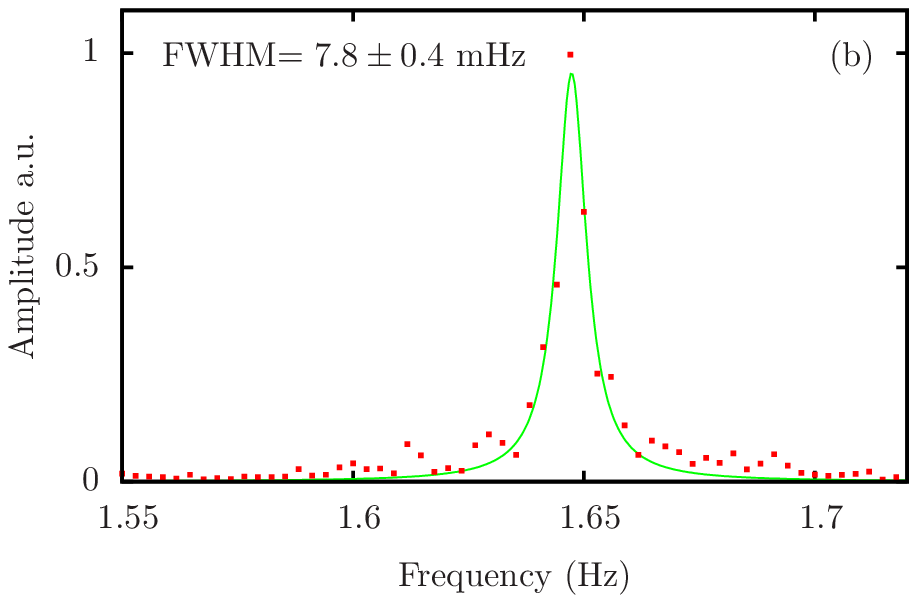}
  \end{center}
  \caption{(a) the Allan deviation for the low-drift data of the time interval 200 s $-$ 540 s of Fig. \ref{fig:result1} (a). (b) the amplitude spectrum of the corresponding waveform. The green line is a Lorentzian fit to the experimental data (red points).}
  \label{fig:resuult2}
\end{figure}

Apart from the limit occurring at short integration times, the long-term stability is also limited and degraded due to drift of the beat frequency. This causes the appearance of a minimum in the Allan deviation. For example, the drift of the beat frequency was occasionally as high as 2 Hz per hour, which in combination with the instability behavior on the short-term, would lead to a (worst-case) minimum of 12 mHz for the Allan deviation. Sometimes, the drift of the beat frequency was very small, e.g., between 200 s $-$ 540 s in Fig. \ref{fig:result1}. In such cases the minimum Allan deviation of the beat \modif{was} as low as $1\times 10^{-17}$ at an integration time of 100 s, as shown in Fig. \ref{fig:resuult2} (a). The corresponding spectrum, obtained by recording the waveform with a DAQ card and performing a FFT, is shown in Fig. \ref{fig:resuult2} (b). The FWHM of the spectrum is $7.8 \pm 0.4$ mHz. This result suggests that if there was no drift, the Allan deviation of the beat signal would keep decreasing to sub-mHz level according to the dependence 17.4 mHz ($\tau/1$ s)$^{-1/2}$.
    
Several causes can contribute to the frequency drift of the beat signal, in particular the temperature drift of the cavity, the drift of the frequency modulation signal, the drift of the laser powers (in this work, only the total power \modif{was} stabilized, not the individual powers). For example, if the coefficient of thermal expansion of the cavity was $1\times 10^{-8}$/K due to a small offset from the optimum temperature, the frequency sensitivity would be about 17.8 Hz/K. For the two laser systems we measured respective sensitivities of 29 mHz and 90 mHz per 1\% change in the laser powers, 1.4 mHz and 2.9 mHz per 1 Hz change in the modulation frequencies, and 200 mHz and 330 mHz per 1\% change in modulation amplitude. Since the above parameters were controlled at much better level, this could not be a reason of the observed drift rate. Furthermore, we do not observe a clear correlation between the room temperature and the beat frequency; the change is less than 0.2 Hz per 2.5 K temperature change. 

The RAM of the laser was also characterized. The modulation strength is about $2\times 10^{-4}$ without use of an optical fiber and about $6\times 10^{-4}$ with the optical fiber in the beam path. The stability of the RAM strength was measured by monitoring the output of the mixer of the PDH circuit when the laser is off resonance. It was found to be similar with and without the optical fiber. The estimated frequency instability caused by the RAM and the mixer is indicated in Fig. \ref{fig:result1} (b), as the black line. Our frequency stability at medium integration time (ca. 100 s) appears limited by the instability of the RAM and might be improved by stabilizing the RAM strength \cite{Muller:03}. On very long time scales, RAM contributes less than 0.1 Hz at $10^{4}$ s integration time, while the variation of beat signal is typically of order 1 Hz. In fact, a long-term measurement of the beat signal over several days shows that the beat frequency oscillates slowly with a peak-to-peak of about 8 Hz, and a period of approx. 10 h. This behavior suggests that the drift of the beat frequency may be caused by residual interferences in the optical path, probably in the optical fiber. 

In summary, we reported on a high-performance laser frequency lock system. By using an approach that avoids the influence of thermal noise of the cavity, we could characterize the lock performance at the mHz level. The locks operated at the intensity noise limits of the lasers and reached a minimum instability of 6.3 mHz ($1.6\times 10^{-17}$) at 40 s integration time (referred to a single lock system). The level is limited by slow beat frequency drift, as evidenced by the observation of instability levels close to $1\times 10^{-17}$ for integration times of 100 s during periods of particularly low drift. The results achieved are an important ingredient for the further development of optical clock lasers and experiments testing Lorentz Invariance using optical resonators. We finally note that the instability of the beat frequency (at 1.78 GHz) is $3.5\times 10^{-12}$. This low relative level might be further reduced by increasing the difference in mode numbers between the two resonator modes used for locking, and could find application for the generation of spectrally pure microwaves. 

This work was performed in the framework of project AO/1-5902/09/D/JR of the European Space Agency.  

\begin{thebibliography}{10}
\bibitem{deltaf}
  \moda{Consider an optical path of length $L$ which changes by $\Delta L$ over a time $\tau$. The change of the phase of a light wave over the time $\tau$ corresponds to a frequency shift $\Delta \nu$.   For a single laser, the \modif{frequency change} is $\Delta\nu =\Delta L\nu/(c\tau)$. \modif{The change of the beat frequency of two lasers is} $\Delta f = \Delta\nu_1-\Delta\nu_2$. In the case of independent optical paths, the beat frequency instability is 
equal to the square root of the sum of squares of the frequency instabilities in the two optical paths, approximately $\sqrt{2}\Delta L \nu/(c\tau)$, if the paths instabilities are similar, $\Delta L_{1} \approx \Delta L_{2} \equiv \Delta L$.
However, in the case of identical optical paths, because $\Delta L_1=\Delta L_2$, $\Delta f = \Delta L (\nu_1-\nu_2)/(c\tau)$, a value smaller by approx. $f/\sqrt{2} \nu$.}
\end{thebibliography}

\end{document}